\documentclass[prl,twocolumn,showpacs,superscriptaddress]{revtex4}
\usepackage{graphicx,epsfig}
\usepackage{times}
\usepackage{graphics,dcolumn,bm,float}
\usepackage{amssymb,amsmath,rotate,color}

\begin{document}
\unitlength 1 cm
\newcommand{\be}{\begin{equation}}
\newcommand{\ee}{\end{equation}}
\newcommand{\bearr}{\begin{eqnarray}}
\newcommand{\eearr}{\end{eqnarray}}
\newcommand{\nn}{\nonumber}
\newcommand{\vk}{\vec k}
\newcommand{\vp}{\vec p}
\newcommand{\vq}{\vec q}
\newcommand{\vkp}{\vec {k'}}
\newcommand{\vpp}{\vec {p'}}
\newcommand{\vqp}{\vec {q'}}
\newcommand{\bk}{{\bf k}}
\newcommand{\bp}{{\bf p}}
\newcommand{\bq}{{\bf q}}
\newcommand{\br}{{\bf r}}
\newcommand{\up}{\uparrow}
\newcommand{\down}{\downarrow}
\newcommand{\fns}{\footnotesize}
\newcommand{\ns}{\normalsize}
\newcommand{\cdag}{c^{\dagger}}

\title{Short range Coulomb correlations render massive Dirac fermions massless}
\author{M. Ebrahimkhas}
\affiliation{Department of Science, Tarbiat Modares University, Tehran, Iran}
\affiliation{Department of Science, Islamic Azad University, Mahabad branch, Mahabad 59135, Iran}
\author{S. A. Jafari}
\affiliation{Department of Physics, Sharif University of Technology, Tehran 11155-9161, Iran}
\affiliation{School of Physics, Institute for Research in Fundamental Sciences (IPM), Tehran 19395-5531, Iran}

\begin{abstract}
Tight binding electrons on a honeycomb lattice are described by 
an effective Dirac theory at low energies. Lowering symmetry by 
an alternate ionic potential ($\Delta$) generates a single-particle gap in the spectrum.
We employ the dynamical mean field theory (DMFT) technique, to
study the effect of on-site electron correlation ($U$) on massive Dirac 
fermions. For a fixed mass parameter $\Delta$, we find that beyond
a critical value $U_{c1}(\Delta)$ massive Dirac fermions become massless.
Further increasing $U$ beyond $U_{c2}(\Delta)$, there will be another phase transition 
to the Mott insulating state. Therefore the competition between the single-particle gap
parameter, $\Delta$, and the Hubbard $U$ restores the semi-metallic
nature of the parent Hamiltonian. The width of the intermediate semi-metallic 
regime shrinks by increasing the ionic potential. However, at small values of $\Delta$, there is a 
wide interval of $U$ values for which the system remains semi-metal.
\end{abstract}
\maketitle

\section{Introduction}
The single-particle spectrum of excitations in graphene~\cite{Geim} can be very accurately
described by a tight-binding model involving hopping between the localized
$p_z$ orbitals of neighboring carbon atoms~\cite{NetoRMP}. The low-energy sector of 
such Hamiltonian will be described by a 2+1 dimensional Dirac theory~\cite{Semenoff}. 
Starting from this point, many perturbations can be imagined and/or fabricated 
to modify the spectrum of Dirac electrons~\cite{Kim}. From applications point of view,
it is important to open up a gap in the spectrum by lowering the symmetry
of the nearest neighbor tight-binding Hamiltonian. For example the substrate can
induce a sub-lattice symmetry breaking, e.g. by adding an alternate
ionic potential of strength $\Delta$, which immediately leads to 
a charge gap of magnitude $2\Delta$ in the spectrum of single-particle
excitations and renders the Dirac electrons massive~\cite{Pyatkovskiy,Semenoff2} the ground 
state of which will be a trivial band insulator (BI). Recent {\em ab-initio} estimates
of the strength of the Hubbard $U$ in graphene suggests that the
on-site Coulomb repulsion is quite remarkable, $\sim 10$ eV~\cite{Wehling}.
Since this value is expected to be 
a local property of $p_z$ orbitals in carbon atoms, one do not expect
the above value of $U$ to be much different when the Dirac fermions
of the underlying honeycomb lattice acquire a mass due to extrinsic
effects, such as substrate, binding with ad-atoms such as hydrogen~\cite{Sofo}, etc. 
Therefore it is important to consider
the competition between the single particle gap parameter $\Delta$ and
the Hubbard parameter $U$ on top of a semi-metallic state at half-filling.
On the strongly correlated limit, when the Hubbard energy scale dominates over 
the ionic potential, i.e. $U\gg\Delta$, the system will be a Mott insulator (MI) 
where the charge fluctuations become costly because of no-double-occupancy
constraint imposed by large $U$, while in the opposite limit
where $\Delta$ dominates over $U$, the system can be described in terms
of an effective massive Dirac theory. The purpose of this paper is to
show show that short range many-body Coulomb interaction $U$ can transform
massive Dirac fermions into massless ones. 


The threefold coordination of carbon atoms on the honeycomb lattice
of graphene is the basic mathematical reason for emergence of cone-like
dispersion in graphene which is responsible for  semi-metallic properties. 
If the dimension and/or coordination number were
different, the parent one-band tight-binding Hamiltonian would 
at half-filling describe a metal. Previous DMFT study by Garg and coworkers 
predicts that for metallic parent Hamiltonians, the intermediate phase will be
a metal~\cite{Garg}, i.e. the competition between Hubbard and ionic terms in
a metal, restores the metallic state. One may ask the
same question for semi-metallic (Dirac) systems: What would be
the result of competition between the Hubbard and ionic terms
in a semi-metal? Would the intermediate phase be a metal?
Or the competition between the Hubbard and ionic terms --both of which
would individually drive the system towards insulating behavior --
will give way for the semi-metallic behavior?

To tackle this question, the dynamical mean field theory (DMFT) 
is a suitable and powerful method~\cite{Georges}, which is capable
of handling Hubbard interaction in a proper way. This approximate technique 
becomes exact in the limit of infinite coordination numbers~\cite{Georges,Caffarel}
which can be formally incorporated by assumption of a higher dimensional
generalization of various lattices~\cite{Georges,Santoro}.
For lower coordination number, the local (k-independent) self-energy employed in the 
DMFT becomes only an approximate description. Hence, the most significant 
drawback of the method is expected to be the underestimation of the 
spatial quantum fluctuations~\cite{Jafari,Ebrahimkhas,Tran}. Therefore the values of the critical
parameters obtained from a simple DMFT on the honeycomb geometry maybe
overestimated~\cite{Wu}. However, the overall picture emerging from this numerically 
powerful method is expected to hold even when more sophisticated methods,
such as cluster DMFT~\cite{Liebsch}, Gutzwiller projection~\cite{Martelo} alternative techniques such as quantum
Monte Carlo~\cite{Sorella} or slave-particle~\cite{Vaezi} schemes are employed. 
Since in this work we are not
interested in fine spectral details, the lack of k-resolution we believe
will not affect the central conclusion of the paper, namely that the 
semi-metallic behavior is restored when Hubbard and ionic potential compete.

The ionic-Hubbard model has been studied extensively in one dimension~\cite{Egami}, 
and two dimensions, and the nature of interim phase has been debated~\cite{Bouadim}. 
In addition to various low-dimensional techniques employed to study such model,
the DMFT technique has also been employed to study the nature of intermediate phase
of this model for metallic parent systems~\cite{Garg, Craco}. In Ref.~\cite{Garg} the authors 
implemented DMFT for the Bethe lattice and found that by increasing $U$ for fixed 
$\Delta$, the system undergoes two phase transition: First from band insulator into 
metallic phase, and second  from metal into Mott insulator phase~\cite{Garg}. 
However in Ref.~\cite{Craco} they used almost the same method and find coexistence 
insulator phase region for some $\Delta, U$~\cite{Craco}.

\section{Model and Method}
The ionic-Hubbard Hamiltonian on the honeycomb lattice with two atoms in same unit cell 
is given by,
\bearr
&& H=-t \sum_{i\in A, j\in B}[c^\dagger_{i,\sigma}c_{j,\sigma}+ h.c.]+ \\ \nn
&& \Delta \sum_{i\in A}n_i -\Delta \sum_{i\in B} n_i + U\sum_{i} n_{i\uparrow}n_{i\downarrow} - \mu \sum_{i} n_i,
\label{IHH}
\eearr
where $t$ is the nearest neighbor hopping, $\Delta$ denotes the ionic potential which 
alternates sign between site in sub-lattice $A$ or $B$. The Hubbard $U$ stands for the 
repulsion potential between two electrons with opposite spins in $i$th site of the lattice.
The chemical potential is $\mu=U/2$ at half filling and so the average filling factor
is $\frac{\langle n_A\rangle + \langle n_B \rangle}{2}=1$.
The model for $t>0$ and noninteracting limit $U=0$, represents a band insulator,
with energy gap $E_{gap}=2\Delta$. The average filling factors at half-filling will become 
$\langle n_A\rangle=0, \langle n_B\rangle=2$. For $U\gg\Delta$, the system is in the Mott insulator 
phase with $n_{A}=n_{B} =1$~\cite{Hafez}. In the intermediate region, a flow-equation analysis
shows that at some energy scale, the ionic and Hubbard terms are expected to cancel 
each other's effect~\cite{Hafez}. From this angle we expect the character of the
parent Hamiltonian (i.e. the hopping term only) to dominate the nature of the
ground state. Therefore for metallic parent Hamiltonians we expect the intermediate
phase to be a metal~\cite{Garg, Hafez}, while for semi-metallic parent Hamiltonian
as will be shown in the following, the semi-metallic character 
will be restored and the massive Dirac fermions will become massless. 
as a result of increasing the Hubbard $U$. 

To formulate the DMFT machinery for semi-metallic parent system, consider a parent tight-binding
Hamiltonian on the honeycomb lattice to be in the paramagnetic phase. 
The interaction Green's function in the bipartite lattice acquires a matrix form as,
\be
G(\vk, \omega^+) =
\left( {\begin{array}{cc}
 \zeta_A(\vk, \omega^+) & -\epsilon(\vk)  \\
 -\epsilon(\vk) & \zeta_B(\vk,\omega^+)  \\
 \end{array} } \right)
 \label{Matrix-G}
\ee
where $\vk$ is the momentum vector in first Brillouin zone,
$\epsilon(\vk)$ is the energy dispersion for the honeycomb lattice~\cite{NetoRMP}, 
and $\zeta_{A(B)}=\omega^+ \mp \Delta +\mu- \Sigma_{A(B)}(\omega^+)$ 
with $\omega^+=\omega+i0^+$. In DMFT approximation the self-energy, $\Sigma_{\alpha}(\omega^+)$ 
is local~\cite{Georges}, i.e. the self-energy matrix is diagonal and independent of $\vk$. Therefore the
off-diagonal elements vanish. Hence the local Green's function corresponding to sub-lattice 
$\alpha=A,B$ can be written as, $G_{\alpha}(\omega^+)=\sum_{\vk} G_{\alpha \alpha}(\vk,\omega^+)$ 
which simplifies to~\cite{Garg},
\be 
G_{\alpha}(\omega^+)=\zeta_{\bar \alpha}(\omega^+) \int^{-\infty}_{\infty} d\epsilon \frac{\rho_{0}(\epsilon)}{\zeta_A(\omega^+)\zeta_B(\omega^+)-\epsilon^2}
\label{local-G}
\ee
where $\alpha=A(B)$, $\bar \alpha=B(A)$ and $\rho_0(\epsilon)$ is the bare DOS of the 
honeycomb lattice~\cite{NetoRMP}. 

We start with an initial guess for the filling factor and self-energy~\cite{Garg}.
Then we determine host Green's function from the Dyson's equation 
$\mathcal{G}^{-1}_{0\alpha}(\omega^+)=G^{-1}_{\alpha}(\omega^+)+\Sigma_{\alpha}(\omega^+)$.
Afterwards, we solve impurity problem and find 
$\Sigma_\alpha(\omega^+)=\Sigma_{\alpha}[\mathcal{G}_{0\alpha}(\omega^+)]$. In this step
we use iterated perturbation theory (IPT) as impurity solver~\cite{Georges}. 
The iteration of these steps continues until convergence is reached.
After convergence, we calculate  the density of states given by 
$\rho_{\alpha}(\omega)=-\sum_{\vk} \mbox{Im Tr}[G_{\alpha}(\vk,\omega^+)]/\pi$. 
The particle-hole symmetry at half-filling leads to $\rho_{A}(\omega)=\rho_{B}(-\omega)$, 
for the DOS of the two sub-lattices. The total DOS for the honeycomb lattice is 
eventually obtained via $\rho(\omega)=\rho_{A}(\omega)+\rho_{B}(\omega)$.

\section{Results and discussion}
\begin{figure}[tbh]
\begin{center}
\includegraphics[width=8cm,angle=0]{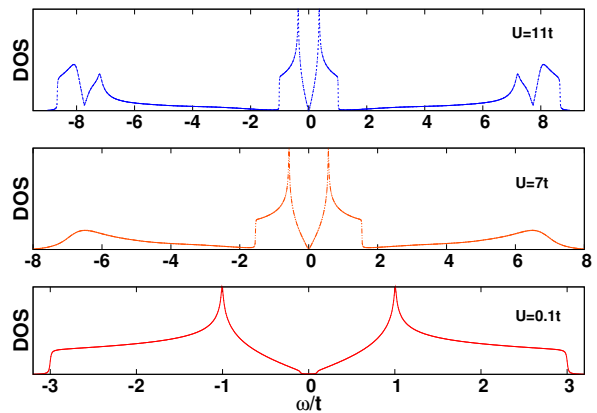}
\vspace{-3mm}
\caption{(Color online) The density of state for honeycomb lattice in ionic-Hubbard model is plotted
for $U=0.1t,7t,11t$ and $\Delta=0.1t$ from down to up. Between band and Mott insulator phase the semi-metal
phase exist as the middle phase. }
\label{DOS.fig}
\end{center}
\vspace{-3mm}
\end{figure}  

In Fig.~\ref{DOS.fig}, we have shown the converged results for three typical values
of $U$ at a constant $\Delta=0.1 t$. As can be seen in this overall picture which covers
the whole range of energies in the bandwidth, for small values of $U$, we have simple
band gap in the spectrum which has single-particle character. As $U$ increases, 
spectral weight is transferred to higher energies to form upper and lower Hubbard
bands. The striking feature in Fig.~\ref{DOS.fig} is that, despite the formation
of Hubbard bands at larger values of $U$, the overall shape of the low-energy 
spectral features resemble that of the parent Hamiltonian. The DOS around the
Fermi level is V shaped at energy scales above $\sim \Delta$, and the van-Hov singularity
arising due to the saddle point in the band structure moves towards lower energies.

\begin{figure}[tb]
\begin{center}
\includegraphics[width=8cm,angle=0]{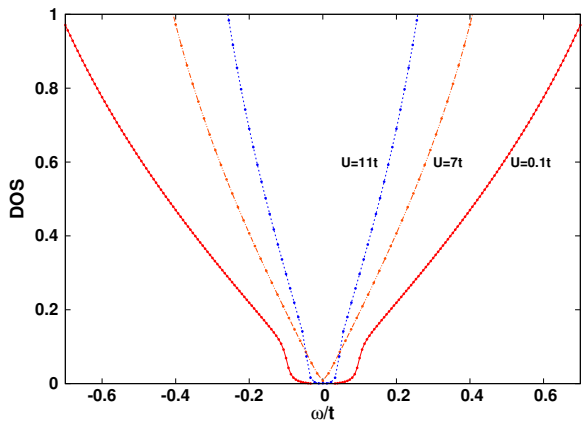}
\vspace{-3mm}
\caption{(Color online) The zoom in the low energy part of the DOS for $U=0.1t,7t,11t$ 
at fixed $\Delta=0.1t$. By increasing $U$, the single particle gap closes to restore
the semi-metallic phase, which is then followed by a SMIT.}
\label{DOS-zoom.fig}
\end{center}
\vspace{-3mm}
\end{figure} 

\begin{figure}[b]
\begin{center}
\includegraphics[width=8cm,angle=0]{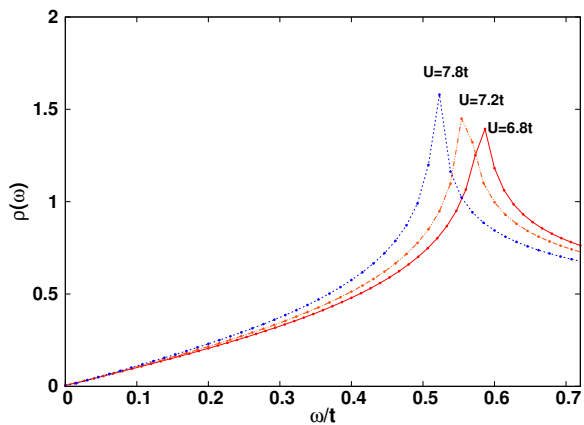}
\vspace{-3mm}
\caption{(Color online) By increasing $U$ the slope of DOS in the intermediate SM region 
changes. This in turn means an increase in the Fermi velocity of electrons. 
This figure corresponds to $\Delta=0.1t$. }
\label{Z.fig}
\end{center}
\vspace{-3mm}
\end{figure} 

In Fig.~\ref{DOS-zoom.fig} we have zoomed in the lowest energy scales around the
Fermi level. As can be clearly seen in this figure, increasing $U$ eventually 
closes the single-particle gap, and restores the linear DOS at all low-energy 
scales, including those below $\sim \Delta$, and renders the system semi-metallic.
To our knowledge, {\em this is a rare instant where many-body interactions of relativistic
fermions effectively removes their mass and makes them massless.}
By further increasing $U$, up to $11t$, the spectral transfer to upper and lower
Hubbard bands becomes more and more (Fig.~\ref{DOS.fig}), while at the same time
a Mott-Hubbard gap appears around the Fermi surface as in Fig.~\ref{DOS-zoom.fig}.
Therefore at a fixed value of $\Delta$, as one increases $U$, there will be a phase
transition at $U_{c1}$ to the parent semi-metallic phase, followed by another
phase transition at larger $U_{c2}$ to the Mott insulating phase.
The intermediate phase is characterized by a strictly linear DOS in the low-energy scales,
which qualifies the intermediate phase for an effective Dirac theory. Note that 
the Fermi velocity in the intermediate effective 2+1 Dirac theory is
renormalized with respect to the effective Dirac theory of the parent Hamiltonian.
In Fig.~\ref{Z.fig}, we show converged results for three typical $U$ values in the
intermediate regime which indicate the decrease in the Fermi velocity of the 
intermediate Dirac liquid phase as a function of $U$ which ultimately leads to
a SMIT~\cite{Jafari}.

\begin{figure}[tb]
\begin{center}
\includegraphics[width=8cm,angle=0]{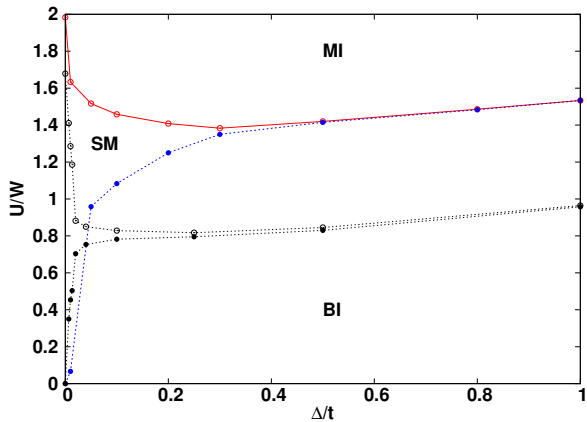}
\vspace{-3mm}
\caption{(Color online) The DMFT phase diagram of the ionic-Hubbard model on the 
honeycomb lattice. Band insulating phase (BI) and Mott insulating phase (MI) are separated
by a semi-metallic phase (SM). For comparison we have also given the similar phase
diagram obtained for metallic parent Hamiltonian~\cite{Garg}. The Hubbard energy scales
in both plots are reported in units of bandwidth $W$.}
\label{phase-diagram.fig}
\end{center}
\vspace{-3mm}
\end{figure}

Repeating such calculations for a range of ionic potential $\Delta$, we map out the 
phase diagram for the ionic Hubbard model on the honeycomb lattice, the boundaries
of which are represented by blue and red lines in Fig.~\ref{phase-diagram.fig}.
First of all for $\Delta\approx 0$, the width of the band insulating (BI) phase 
diminishes which is equivalent to $U_{c1}\approx 0$, and there will be only
one phase transition separating SM and MI phases~\cite{Jafari,Ebrahimkhas,Tran}.
Note that the vertical axis denotes values of $U/W$, where $W$ is the bandwidth
of the parent Hamiltonian. As can be seen in Fig.~\ref{phase-diagram.fig}, 
by increasing $\Delta$, the SM region becomes narrower, and beyond $\Delta/ t\sim 0.5$
-- the energy scale up to which the Dirac cone linearization holds --
the red and blue boundaries merge. Beyond this point, the phase transition
will be essentially between the BI and MI phases.

For comparison in Fig.~\ref{phase-diagram.fig} we have also reproduced the
data from Ref.~\cite{Garg} for a metallic parent Hamiltonian with Bethe lattice
DOS. One interesting difference between the intermediate region on the honeycomb
and Bethe lattice is that the in the former case, the ground state of the
parent Hamiltonian (SM) occupies a much larger region in the two dimensional space
of $(\frac{\Delta}{t},\frac{U}{W})$. This could be interpreted as the robustness
of the parent phase of the Dirac electrons compared to Schroedinger electrons
when an ionic and Hubbard energy scales compete. 
  One more conclusion could be drawn from Fig.~\ref{phase-diagram.fig} is that
if we imagine a horizontal line at a constant $U_{g}/W\sim 0.5-0.7$ which maybe
relevant to graphene~\cite{Wehling}, it can be seen that for a range of $\Delta/t\sim 0.04-0.05$,
the system still remains in the SM phase. For a reasonable estimate of $t\sim 2.7-3$ eV, 
this implies that, graphene can remain SM despite a symmetry breaking
ionic potential of strength $\Delta\sim 110-150$ meV. Such a single-site DMFT
may have overestimated the upper boundary $U_{c2}(\Delta)$~\cite{Wu}. Improving on
this calculations by e.g. cluster extension of DMFT is expected to push the 
upper boundary down, and hence our estimate of the tolerance $\Delta\sim 110-150$ meV
is not expected to change much.

\begin{figure}[tb]
\begin{center}
\includegraphics[width=8cm,angle=0]{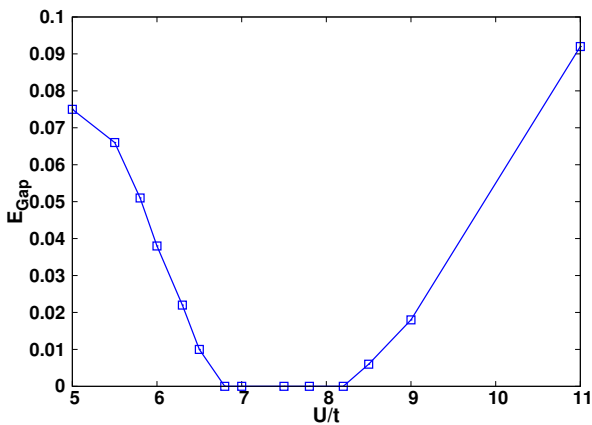}
\vspace{-3mm}
\caption{(Color online) The $E_{\rm Gap}$ in units of $t$ as a function 
of $U/t$ for $\Delta=0.1t$ is plotted. Presence of non-zero $U$ strongly 
modifies the actual gap, compared to the "non-interacting gap" $2\Delta$.}
\label{gap.fig}
\end{center}
\vspace{-3mm}
\end{figure} 

Normally the gap magnitudes in gapped graphene samples are assumed to 
satisfy $E^{(0)}_{\rm Gap}=2\Delta$. However this picture holds when
one completely ignores the many-body effects. Hence we have used the
superscript "$0$" to emphasize that it is "non-interacting" gap. 
In Fig.~\ref{gap.fig} we plot the actual spectral gap as a function of $U$ 
for a fixed value of $\Delta=0.1t$. As can be seen the actual (interacting) gap
strongly depends on the interaction parameter $U$ in both band- and Mott-insulating
phases. In particular the value of gap corresponding to $U\sim 4t$ which is
relevant to graphene, at $\Delta=0.1t\sim 280$meV is expected to be less than 
half of the non-interacting estimate. Hence without proper account of the
on-site correlations~\cite{Wehling}, the estimates of $\Delta$ based on the
observation of $E_{\rm Gap}$ is expected to be about a factor of two underestimated.

\section{Summary and discussions}
We studied the ionic-Hubbard model on honeycomb lattice by DMFT method, and we found 
the system under goes two phase transition by increasing $U$ for $\Delta>0$. 
For $U<U_{c1}(\Delta)$ the system is in band insulator phase whose effective
low-energy theory is 2+1D massive Dirac theory. 
For $U_{c1}(\Delta)<U<U_{c2}(\Delta)$ region 
the competition between $U$ and $\Delta$ energy scales restores the semi-metallic character
of the parent Hamiltonian and makes them massless Dirac fermions.
Ultimately, for $U>U_{c2}(\Delta)$ the strong correlations transform the 
intermediate SM to a Mott insulator~\cite{Jafari}. 
Beyond $\Delta\sim 0.5t$, the upper and lower critical values
approach each other and the semi-metallic region will be so narrow that it will disappear. 
This theoretical study has implications for gapped graphene samples. The observed values
of spectral gap is always below the "non-interacting" value of $2\Delta$. 
The strong re normalization of the gap magnitude is not the only manifestation of
the Hubbard correlations $U$. It also strongly influences other spectral features
such as the life-time of quasi particles, e.g. in hydrogenated graphene~\cite{DannyPSSB}.

\section{Acknowledgements}
The authors are grateful to  A. Garg for useful discussions.
S.A.J. was supported the National Elite Foundation (NEF) of Iran.

\end{document}